# Optomechanics of chiral dielectric metasurfaces

Simone Zanotto[1], Alessandro Tredicucci[1,2], Daniel Navarro-Urrios[3], Marco Cecchini[1], Giorgio Biasiol[4], Davide Mencarelli[5], Luca Pierantoni[5], Alessandro Pitanti[1]

[1] NEST, CNR Istituto Nanoscienze and Scuola Normale Superiore, piazza San Silvestro 12, 56127 Pisa - Italy

[2] Dipartimento di Fisica, Università di Pisa, largo B. Pontecorvo 3, 56127 Pisa – Italy

[3] MIND-IN2UB, Departament d'Enginyeria Electrònica i Biomèdica, Facultat de Física, Universitat de Barcelona, Martí i Franquès 1, 08028 Barcelona, Spain

[4] Istituto Officina dei Materiali CNR, Laboratorio TASC, Basovizza (TS) - Italy

[5] Università politecnica delle Marche, Ancona 60131, Italy

**Electromagnetic fields coupled with mechanical degrees of freedom have recently shown exceptional and innovative applications, ultimately leading to mesoscopic optomechanical devices operating in the quantum regime of motion. Simultaneously, micromechanical elements have provided new ways to enhance and manipulate the optical properties of passive photonic elements. Following this concept, in this article we show how combining a chiral metasurface with a GaAs suspended micromembrane can offer new scenarios for controlling the polarization state of near-infrared light beams. Starting from the uncommon properties of chiral metasurface to statically realize target polarization states and circular and linear dichroism, we report mechanically induced, ~300 kHz polarization modulation, which favorably compares, in terms of speed, with liquid-crystals commercial devices. Moreover, we demonstrate how the mechanical resonance can be non-trivially affected by the input light polarization (and chiral state) via a thermoelastic effect triggered by intracavity photons. This work inaugurates the field of Polarization Optomechanics, which could pave the way to fast polarimetric devices, polarization modulators and dynamically tunable chiral state generators and detectors, as well as giving access to new form of polarization nonlinearities and control.**

Rapid technological advancements in micro- and nano-fabrication have recently lead to a class of photonic resonators strongly intertwined with mechanical elements[1]. This new field - Cavity Optomechanics - has lead the research efforts in photonic solid state devices showing exceptional properties, which ultimately resulted in the ground state cooling of motional modes in a mescoscopic system through radiation pressure[2]. Not limited to this milestone result, a proper assessment of the mechanical motion in optomechanical systems can be used to strongly influence the output light, not only at a quantum level, producing squeezing[3,4] and non-classical states[5], but even in a classic picture, with chaotic light[6] and low-power, fast intensity modulation[7–9].

On the other hand, the field of photonic artificial materials has been a strong player in the field of electromagnetic structured media for more than two decades[10–12]. Wavelength-scale patterns of scatterers have shown the capability to produce bright colors and extravagant surface effects, often mocking what elegantly realized in biological systems[13]. In particular, nanostructuration techniques combined with powerful *ab initio* design tools allow to develop materials which show a chiral response far larger than their natural counterparts, i.e., stereochemical compounds.

In this article we report how we can add more functionalities to an all dielectric chiral metasurface by defining its own pattern on a mechanically actuable GaAs membrane. The fundamental vibrational mode of the membrane produces a modulation of a complex combination of light intensity and/or polarization, as will be detailed later. Conversely, the polarization state of the photons addressing the device and, particularly relevant, their chirality, can deeply influence the mechanical resonator by shifting its resonant frequency by a thermoelastic effect induced optical spring. The fast dynamical frequency, exceeding 300 kHz, makes this new kind of device appealing for possible future polarization modulators or fast polarimeters. Furthermore, while mechanics and temperature have somewhat been employed before for statically reconfigurable metasurfaces[14–17] to the best of our knowledge, our system is the first to employ the mechanical motion for a dynamical control of the optomechanical device. Also, all mechanically reconfigurable metasurface realizations are based on metallic scatterers, which offer degraded performances due to ohmic losses with respect to the all-dielectric system we report here.

The fabrication of our device starts with patterning a periodic array of holes on a 220 nm thick GaAs layer. Gallium arsenide material has several advantages: it is easily machinable[18], it can host active elements[19–21], and it has strong nonlinear optical response[10,22–25]. Moreover, being its refractive index similar to that of silicon, the design presented here can be straightforwardly exported to CMOS-compatible platforms for further integrability with electronics. Few GaAs based optomechanical devices have also been reported[9,26,27], yet not including chiral metasurfaces. The chosen pattern design is a minimal one (see the SEM micrograph of fig. 1g) and it is "L-shaped"; this is the simplest shape that breaks in-plane mirror symmetry, granting for 2D-chirality. When the patterned membrane is considered in conjunction with the underneath substrate, the overall structure becomes truly 3D-chiral. The photonic response of the structure is thus a combination of two effects: the resonance of the sole patterned membrane[28] and the multiple Fabry-Pérot resonances determined by membrane-substrate optical paths, as sketched in Fig. 1b. Due to interference, the overall device optical response depends on the relative membrane to substrate distance; this gives a dissipative form of optomechanical coupling which we exploit in our experiment. The membrane vibration impacts on the phase, intensity and polarization of light and therefore can be investigated by looking at the device optical response. This is done using a standard free space setup where the polarization state of input light can be easily controlled, as shown in Fig. 1a. In most of the experiments, membrane oscillations are forced by a piezoelectric actuator, and the coherent response is detected through a lock-in amplifier. The mechanical mode we use is the fundamental one, which we independently characterized using a Laser Doppler Vibrometer (see Fig. 1e). The simulated mode (predicted at 0.34 MHz according to full 3D FEM simulations, see Fig.1f) well reproduce the displacement field evaluated in the interferometric measurement.

The mechanical effect on the optical response of the device can be easily described by means of reflection matrices, which describe the amplitude, phase and polarization response of the metasurface-substrate complex. The reflection matrix $R$ connects the incident and the reflected wave field components by $\mathbf{E}_r = R\,\mathbf{E}_i$, being $\mathbf{E}_{i,\,r} = (E_{i,r;\,x}, E_{i,r;\,y})$ the Cartesian components of the complex electric-field vector associated to the propagating waves. In addition to the fixed metasurface geometric parameters, the $R$ matrix depends on the metasurface-substrate distance $\zeta$; we assume that the focused laser beam, smaller than the membrane size, locally probes rigid membrane shifts, leading to approximating $\zeta$ as a constant number, independent on the position across the membrane. We can hence write, for small oscillation amplitudes,

$$R = R(\zeta) = R_0 + R_1 \Delta\zeta$$

where $R_1 = \partial R/\partial \zeta$ and $\Delta\zeta$ is the displacement from the rest position. From this, static and dynamic response of the metasurface can be easily determined.

Static response follows from reflection coefficients encoded in $R_0$. As an example, experimental cross-polarized (Horizontally-polarized input, Vertically-analyzed output) and circular dichroism (difference between output intensities for right-circular and left-circular inputs) reflection spectra are reported in Fig. 1c, where the Fabry-Pérot fringes have been artificially removed for the sake of clarity. Our formalism well reproduces the measured spectra which mainly consist of a broad resonant feature centered at about 1545 nm. The simulated field profile of the corresponding optical mode is reported in Fig. 1d.

Dynamical response consists in a modulation of phase, amplitude and polarization state of the reflected wave ($R_1$) around the point dictated by static response ($R_0$). Particular forms of static and dynamic reflectivities can be targeted at the design stage; in particular, it has been demonstrated that using an appropriately complex metasurface unit cell, arbitrary control on the reflectivities can be achieved[29]. On the other hand, our device employs a minimal design, resulting from the maximization of circular dichroism at 1550 nm. Despite its simplicity, it offers a rich landscape of static responses in its spectrum of operation (i.e. both positive and negative circular and linear dichroisms) and perfectly illustrates the physics behind static and dynamic control of polarization states. This is shown by performing a modulation experiment where the incident polarization is kept fixed to the horizontal (H) state (Fig 2), and the intensity modulation of the reflected wave is detected. To fully illustrate the metasurface dynamical response, we devised two measurements, with/without a linear polarizer placed in the reflected beam with vertical (V) pass axis, respectively. The results are reported in Fig. 2b, together with the simulations for comparison; the theoretical formalism is provided in the SI. Note that the modulation spectrum shows evident fringes, dictated by multiple Fabry-Pérot resonances; superimposed to it a weak envelope centered around the metasurface resonance can be observed. The measurement without the analyzer indicates that the metasurface oscillation is capable of inducing a pure intensity modulation by acting on the modulus of the reflected field $|\mathbf{E}_r|^2$; by comparing this spectrum with what obtained with the analyzer, an interesting feature appears: the cross-polarized spectrum (dark blue curve) shows a completely different wavelength dependence, evidencing a dynamic polarization state modulation. If the polarization state were only rotated in a static way, the cross-polarized modulation spectrum would be simply a fraction of the spectrum taken without analyzer; as this is not the case, it means that the polarization state itself is modified by the mechanical motion.

To better clarify this aspect we performed further theoretical analysis in the direction of full dynamic polarimetry of the reflected light. Having access to the matrices $R_0$ and $R_1$, it is straightforward to determine the Stokes parameters of the reflected radiation and its dependence on the displacement $\zeta$. A natural metric to quantify the polarization modulation effect is to measure the distance between two points on the Poincaré sphere calculated for $\zeta = 0$ and $\zeta = 0.8$ nm (i.e., the membrane displacement measured in the experiment discussed above). This distance, stated in terms of the angle subtended between the two polarization points and the sphere center, is plotted with its spectral dependence in Fig. 2c, where we highlighted the peak maxima with circular markers. Both the strong Fabry-Pérot oscillations and the envelope effect of of the metasurface resonance are still evident. A maximum polarization state modulation of roughly 0.02 rad/nm can be appreciated. By increasing the piezoelectric actuator driving voltage, we can linearly increase the harmonic modulation value up to about 0.2 rad (when the device operates in vacuum). The system main limitation occurs when the mechanical motion becomes nonlinear, giving rise to an anharmonic time dependence, which in any case can still produce modulation amplitude exceeding 0.5 rad (see SI).

More insights are gained by looking at the actual location of the polarization state on the Poincaré sphere and at the path followed by the polarization point on the sphere when mechanical oscillation occurs. These important features are illustrated in Fig. 2d. Here, we analyzed a selection of wavelengths close to the metasurface resonance (peaks of fig. 2c in the gray shaded area). The colored segments on the sphere represent the polarization modulation occurring when $\zeta$ is tuned from -3 to 3 nm. Such a larger displacement is still experimentally accessible, as discussed, at the expense of losing perfect mechanical linearity. When the modulation is mapped on the Poincaré sphere (Fig. 2d) it illustrates a very interesting behavior: in the metasurface resonance region, *both* static polarization state *and* polarization modulation "direction" can be chosen in a non-trivial way. Here, tuning parameter is wavelength; more generally, static and dynamic polarization response can be chosen by appropriately targeting wavelength and geometric parameters. It can be envisaged that more complex structures, which go beyond the proof-of-principle L-shaped array analyzed here, could lead to designer's targeted polarization and wavelength response.

Linear polarization rotation is a precursor of chirality, in the sense that chiral objects generally rotate the polarization plane of a linearly polarized beam. However, a more direct measure of electromagnetic chirality is circular dichroism (CD). We hence measured the optomechanical effect on circular dichroism by measuring its modulation induced by membrane oscillation. The result is illustrated in Fig. 3: the spectral behavior displays clear fingerprints well reproduced by the full wave electromagnetic simulation, even in their finer details. These data witness two main points. First, the device is a wavelength-sensitive modulator acting on circular polarization, which can replace other technologies like liquid crystal based modulators. The advantage lies in the possibility to avoid any encapsulation technique needed to confine the liquid phase, which hinders miniaturization and integration of the components; moreover, the actuation frequency of our device is larger than that typically achieved in liquid crystal devices (hundreds of kHz versus tens of kHz). Also, despite the narrow operation bandwidth of our prototype, improved designs can be realized to target specific operation wavelengths and modulation frequencies. Second, mechanical oscillations act on the proper, near-field based electromagnetic chirality of the metasurface, eventually connected to the capability that local fields have to excite chiral molecules[30,31]. Since this mechanism can be employed to discriminate between the helicity of analytes, our chirality tuning method may be employed in sensing platforms where the local electromagnetic chirality is required to be modulated at a high frequency.

The phenomena analyzed so far have been studied in ambient conditions. While this demonstrates the robustness of the sample, the presence of air at atmospheric pressure degrades the mechanical quality factor, preventing from a full exploitation of the potentials of the optomechanical interaction. To this end we enclosed the sample in a vacuum chamber and analyzed again the mechanical response. By reducing the air-induced damping, at a base pressure of 3e-5 mbar, the mechanical features can be well fitted with Lorentzian lineshapes (see the quadrature-space fits of Fig. 4b) holding linewidths as narrow as 0.15 kHz, translating in Q-factors of about 2000.

At first, we focused on the light response over a linearly, H-polarized input. The detected signal is reported in Fig. 4a, as a function of laser wavelength and mechanical driving frequency. In the analyzed range, which spans slightly more than one Fabry-Pérot fringe, different levels of modulation amplitude are accompanied by a redshift in the mechanical resonant frequency. Such a shift depends linearly on the optical power, as shown in Fig. 4b-c, and has a peculiar wavelength dependence. Figure 4e shows how the resonant peak amplitudes scale with the optical wavelength; the very different spectral shapes of the signal with and without the vertical analyzer confirms the fact that the device is modulating intensity and polarization in a non-trivial way, as shown in Fig. 2. Note that the simulations (continuous lines) well reproduce the experimental results (dot-connected full markers). More interestingly, the spectral shift of the mechanical resonator is strongly correlated with the intracavity photon energy density, as can be seen in the two-axis plot of Fig. 4f. This effect has a genuinely optomechanical nature and can be understood as the optical spring effect: the presence of electromagnetic field in the structured metasurface modifies the effective Hooke constant of the equivalent mass-spring system. To understand the origin of the mechanical spring effect in our device, it is necessary to go beyond the single resonance analytical model well known in the optomechanics community[1], due to the presence of a multitude of Fabry-Pérot resonances superimposed with the metasurface resonance. We hence performed a numerical study of the forces acting on the membrane, from which it resulted that the forces are of thermal origin. Indeed, the effects of radiation pressure force and of its gradient are ruled out, as they change sign in the relevant wavelength range and because they are far too small to justify the observed shifts (see SI). Moreover, electrostriction effects can be safely neglected even assuming the large photoelastic coefficients of GaAs[32]. Instead, we found an excellent correlation between the intracavity field and the opposite of the frequency shift (Fig. 4f). In essence, a little residual absorption, likely attributable to impurities in the semiconductor, drives a wavelength-dependent heating of the metasurface, with a consequent relaxation of its effective spring constant. The upper bound for temperature rise is estimated to be a fraction of Kelvin, compatible with the simulations of the thermally induced frequency shifts, as detailed in the SI.

Optical spring effect is one of the cornerstones of cavity optomechanics, and one of the main signs of back-action effects; as such it has been observed in several systems[1]. Nonetheless, the interplay between this effect and light polarization has far less been explored; to date, only few theoretical studies can be found on the topic[17]. Our device has proved instead to be a precious test-bed for understanding how the optical forces can be harnessed by means of polarization; moreover, it revealed an intriguing interplay between polarization-dependent optical forces and polarization-dependent optomechanical transduction. The experimental framework is the same as in the previous point, with the difference that here light wavelength is kept fixed at 1545.8 nm and that the polarization state of the probe beam is continuously tuned across the Poincaré sphere. The results are reported in Fig. 5. Panel (a) illustrates that the mechanical transduction curve changes resonance frequency and magnitude as a function of the light polarization state. In particular, the two opposite circular polarization states correspond to extreme values of frequency shift, while the opposite diagonal linear polarization states correspond to extreme levels of optomechanical transduction. Note that here the transduced signal is entirely given by intensity modulation, as no analyzer is placed after the sample. In panel (b) we report the full polarimetric characterization of intensity modulation: each dot corresponds to an input polarization state set in the experimental process, while the gray lines are zero and first order spherical harmonics (more on this later). By comparison, panel (e) illustrates what is expected from the fully-vectorial electromagnetic model (panel (d) provides an alternative representation on the three-dimensional Poincaré sphere). As can be seen, we obtain a good agreement between experiment and simulations. Considering now the spring effect, panel (c) presents the mechanical frequency shift on the full polarization space. As previously observed, the shift (changed in sign) is correlated with the intracavity field [panel (f)], corroborating the picture that the spring effect has a thermo-optic origin. An additional feature of both optomechanical transduction and frequency shift is that they have a dipole-like dependence on the sphere. In other words, the quantities $\Delta I/I_0$ and $\Delta \omega_m$ have contributions from the zero and the first order spherical harmonics, as expected from functions that are bilinear in the polarization Jones vector (see SI). We finally notice that the intensity modulation and the frequency shift have a different – and almost complementary – dependence on the polarization state; as an example, comparing the two polarimetric maps of Fig. 5, we have almost zero circular dichroism modulation (the intensity modulation at the R and L poles are roughly the same), accompanied by a strong dependence of the frequency shift from the chirality of light (zero at the R pole, maximum at the L pole). This means that, in principle, the optomechanical chiral effects can be exploited to implement a polarimeter, based on a completely different mechanism with respect to ordinarily employed devices and with the potential to be orders of magniture faster.

In conclusion, we have demonstrated that the handedness-sensitive optical response of a chiral metasurface can be coupled to nanoscale mechanical motion. When this is done, fast linear and circular polarization modulation is realized; also, the mechanical resonant features are affected by the chirality of a drive light beam. With the potential to be scaled up to high-frequency mechanical mode (~GHz) coherently excited by Surface Acoustic Waves, our device has the potential to strongly impact the field of polarization control. Furthermore, the polarization nonlinearity induced by the nontrivial mechanical interaction can pave the way to complex and fast operation on the polarization state of light, as polarization squeezing. Given the ubiquity of polarized light in fundamental science and in applications, we believe that our proof-of-principle experiment will open new avenues in several fields such as biophotonics, material science, drug discovery, and telecommunications.

**Methods**

***Sample fabrication*** Dielectric metasurface has been fabricated starting from an epitaxial heterostructure constituted of a AlGaAs/GaAs bilayer on a GaAs double-polished wafer. Al concentration in AlGaAs layer is 50%, thicknesses of the layers are, respectively, 1455 and 210 nm as determined by spectroscopic ellipsometry. Few nm of oxide (n ≈ 1.5) are found on the GaAs/air interface. The L-shaped hole array is defined by electron beam lithography (All Resist AR 6200 CSAR resist, exposure with a Zeiss Ultraplus SEM, 30 kV acceleration, pattern generated with Raith ELPHY). Subsequent dry etching (Sentech ICP-RIE reactor, $Cl_2$/$BCl_3$/Ar 6/1/10 sccm, 25W / 100W on HF and ICP plasmas, ≈ 1 nm/s etch rate) and wet etching (concentrated HF solution, 1 min dip) led to the released membrane. Hole sizes are shown in the SI. The array is constituted by 50 equispaced periods.

*Optical modulation experiment* The optical bench follows the scheme reported in Fig. 1. There, a pointing HeNe laser made collinear with the main telecom laser through a dichroic mirror is not illustrated, as well as the vision system (CCD plus zoom lenses) which is also not reported. The waveplates, which in Fig. 1 are placed before the beamsplitter for illustrative purposes, are in reality placed either before or after it depending on the experiment: for the measurements of Fig. 2 and 4, they are placed before the beamsplitter; for those of Fig. 3 and 5 they are placed after; the reason is that the retardance of the beamsplitter was not known. The focusing lens is an achromatic doublet with 75 mm focal length. All the components have been purchased from Thorlabs, with the exception of the tunable laser and the detector (New Focus, respectively, Velocity TLB-6700, 1520-1570nm and IR nanosecond photodetector 1623). Detection is provided by a Zurich Instruments UHF lock-in instrument in the case of coherent measurements, ad by an Agilent 4395A spectrum analyzer for the incoherent ones. Broadband spectra (Figs. 2 and 3) are collected by scanning the laser wavelength and collecting the data with an oscilloscope. Wavelength was monitored by sending the light exiting from the secondary port of the beamsplitter on a Fabry-Pérot etalon constituted by a double-polished bare GaAs wafer.

*Vibrometric measurement* The vibrometric map reported in Fig. 1e has been collected by means of a Polytec Laser Doppler Vibrometer instrument. The quality of the map is slightly degraded due to the presence of holes on the membrane and to the absorption of GaAs at the wavelength of the laser employed by the instrument ($\lambda$=523nm).

*Numerical analysis* Electromagnetic simulations have been performed by a hybrid scattering-transfer matrix method plus rigorous coupled wave analysis, as detailed in SI. MATLAB code has been extensively used, also with built-in optimization algorithms to target the resonance frequency. The electromagnetic module of finite-element COMSOL Multiphysics solver has been employed to produce the map of Fig. 1d. The continuum mechanics module of COMSOL has instead been employed to determine the vibration eigenmodes of the membrane (Fig. 1f).

**Author contributions**

S.Z. and A.P. conceived the experiment and performed the optical measurements. G.B. provided the epitaxial material and S.Z. fabricated the device. D.M. and L.P. supported the numerical analysis. M.C. contributed to the mechanical characterization. S.Z., A.P, D.N.-U. and A.T. analyzed the results. S.Z. and A.P. wrote the manuscript with suggestions and contributions from all the other authors.

**Acknowledgement**

This work was supported by the European Commission project PHENOMEN (H2020-EU-713450). DNU acknowledges the support of a Ramón y Cajal postdoctoral fellowship (RYC-2014-15392). We acknowledge Alessandra Bontempi - CNIT for ellipsometric measurements, and Francesco Riboli and Sara Nocentini – LENS, for preliminary optical characterization of an older device set.

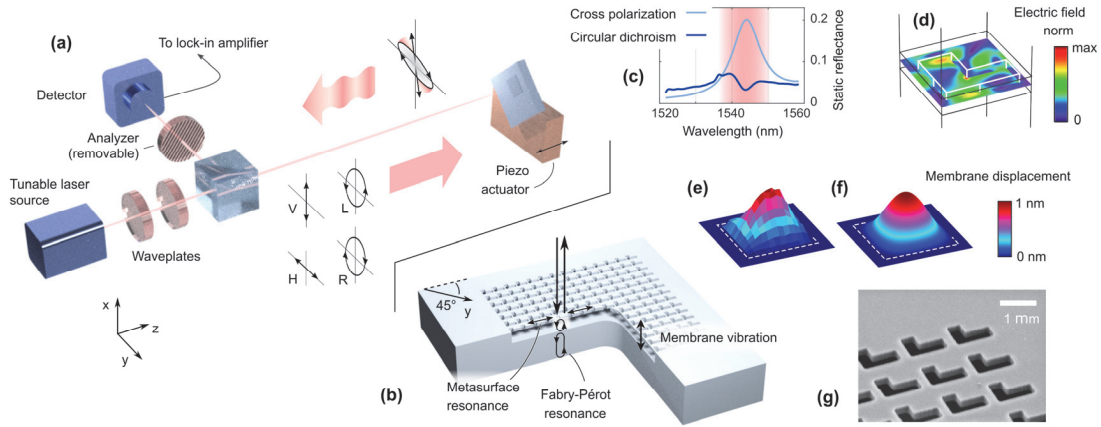

**Figure 1. Optomechanics of a chiral metasurface.** (a) Sketch of the setup employed in the experiment and schematics of the effects observed. An arbitrarily polarized laser beam is reflected off the metasurface, with its intensity and polarization modulated in time. The device can be mounted in a vacuum chamber to enhance the mechanical resonance quality factor. The device is represented in panel (b): a semiconductor (GaAs) membrane is patterned with L-shaped holes, responsible for a chiral resonance arising from the combined metasurface effect and of the Fabry-Pérot effect. The photonic response is sensitive to the membrane vibration, as it affects the Fabry-Pérot length. Chiro-optical effects occur in the wavelength range relevant for telecommunications (c). Here, the Fabry-Pérot effect is filtered out and the cross-polarized (H-polarized input, V-analyzed output) and circular dichroism (difference between reflection spectra with Right/Left circularly polarized inputs) curves are fingerprints of the metasurface resonance only. The field profile of the mode responsible for such effects is plotted in (d). Panels (e) and (f) illustrate the membrane displacement occurring when the fundamental mode is excited, respectively, in an experiment (laser Doppler vibrometry) and in a finite-element model. Panel (g) is a scanning electron micrograph of the metasurface.

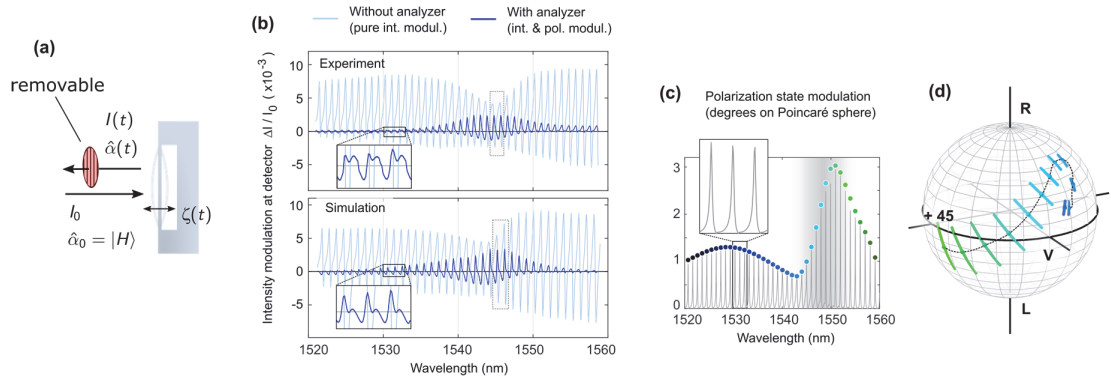

**Figure 2. Optomechanical fingerprints on linear polarization.** (a) Schematic of the experiment. Linearly horizontally polarized light (see also Fig. 1 for the convention) is incident with a constant intensity on the metasurface. As membrane oscillates in time (0.8 nm oscillation amplitude), intensity and polarization are modulated, as reported in (b). The traces show the wavelength-dependent intensity modulation observed at the detector, i.e., after the sample and a possible linear analyzer. Light blue traces are obtained without the analyzer, hence quantifying the pure intensity modulation. Dark blue traces are obtained with the analyzer, thus being contributed from intensity and polarization modulation. The wavelength range highlighted in the gray box is object of further analysis (see Fig. 4). The fast wavelength oscillations are due to Fabry-Pérot effect in the substrate. (c) Calculated polarization modulation spectrum for the same experimental conditions. (d) Polarization rotation and modulation represented on the Poincaré sphere around the metasurface resonance (each colored segment corresponds to the wavelength of a peak of the curve in (c), in the gray-shaded region). The center of each segment is the steady-state output polarization, still for H-polarized input. Oscillation amplitude is here assumed to be 3 nm.

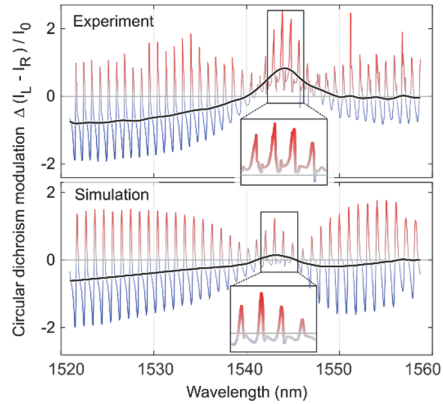

**Figure 3. Optomechanical fingerprints on circular polarization.** Circular dichroism modulation occurring upon mechanical oscillation of the membrane. Narrow wavelength features are due to Fabry-Pérot interference occurring in the substrate, while the envelope shape is dictated by the metasurface resonance.

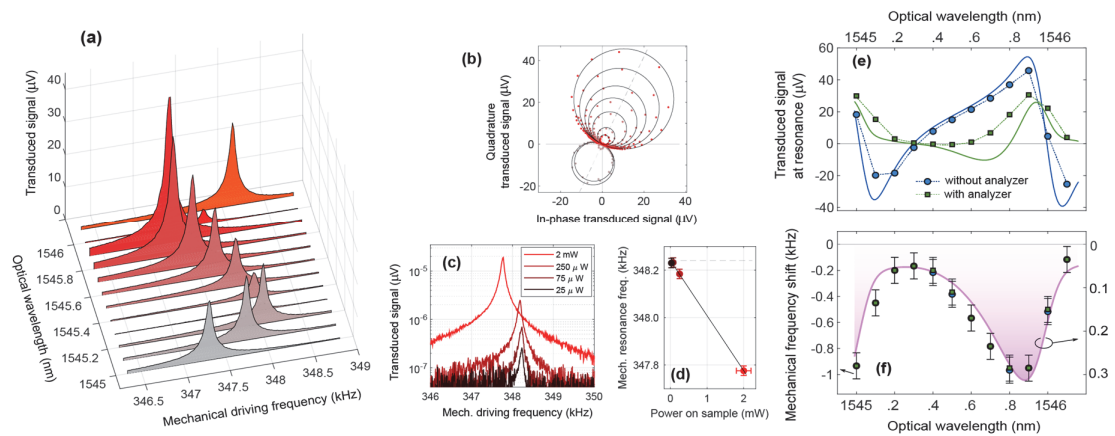

**Figure 4. Optomechanical spring effect in the high mechanical Q-factor regime.** (a) Evolution of the mechanical resonance, in vacuum around one Fabry-Pérot fringe, with both peak value and central frequency dependent on the optical wavelength. Input polarization is H and no analyzer is present. (b) Fit of the transduced signal in the quadrature space. Dependence of the transduced mechanical spectrum upon optical power (c) and resonance red-shift with optical power (d). (e) Optical wavelength dependence of the transduced signal peak across one Fabry-Pérot fringe (region in the gray-shaded area of Fig. 2b). Both measurements with and without an output analyzer (solid symbols) are well reproduced by simulations (solid lines). (f) Wavelength dependence of the spring effect. No dependence on the output polarization filtering is observed, and the data are strictly correlated with the intracavity energy density.

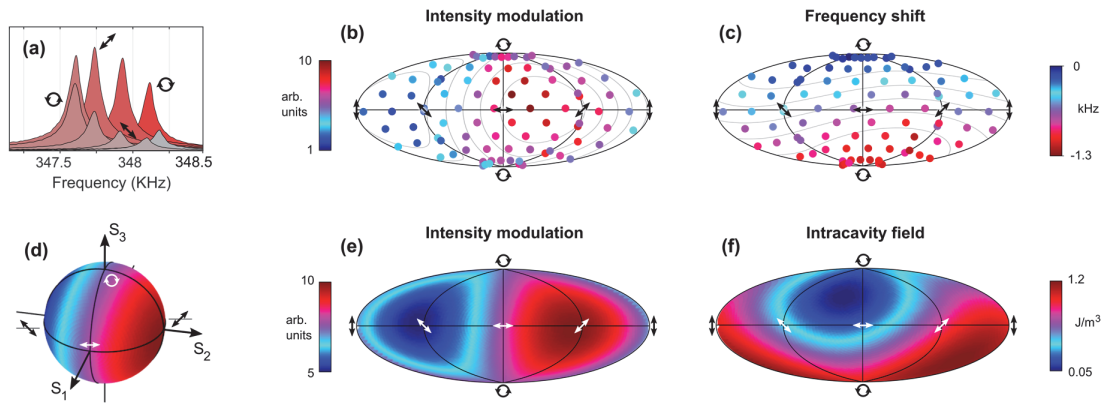

**Figure 5. Observation of chiral spring effect.** (a) Mechanical resonance evolution as a function of the chirality of input light. Full polarimetric analysis of modulation amplitude and frequency shift (b-c) shows a different mapping on the projected Poincaré sphere. Continuous lines are spherical harmonics. The experimental results are validated by the simulations (d-f). While (e)-(f) are plotted in the same configuration as the experimental data, panel (d) offers a different, 3D representation of the results of panel (e). Frequency shift is negatively correlated with the intracavity field, confirming the thermal nature of the chiral spring effect.